\documentclass[showpacs,showkeys]{revtex4}
\usepackage{graphicx}
\def\lsim{\mathrel{\rlap{\lower4pt\hbox{\hskip1pt$\sim$}}
    \raise1pt\hbox{$<$}}}         
\def\gsim{\mathrel{\rlap{\lower4pt\hbox{\hskip1pt$\sim$}}
    \raise1pt\hbox{$>$}}}         

\begin{document}

\title{Solar Neutrinos: Models, Observations, and New Opportunities}

\author{W. C. Haxton}

\affiliation{Institute for Nuclear Theory and Department of Physics, \\
University of Washington, Seattle, WA 98195, USA}

\begin{abstract}
I discuss the development and resolution of the solar neutrino problem, as well as
opportunities now open to us to extend our knowledge of main-sequence
stellar evolution and neutrino astrophysics. 
\end{abstract}

\email{haxton@phys.washington.edu}

\maketitle

\section{Introduction: A Brief History of the Solar Neutrino Problem \cite{jandr}}
This paper is based on a talk given at the Caltech conference \cite{caltech}  ``Nuclear 
Astrophysics 1957-2007: Beyond
the First 50 Years," July 23-27, 2007, which focused on the state of nuclear astrophysics
fifty years after the seminal paper of Burbidge, Burbidge, Fowler, and Hoyle \cite{bbfh}.
The quest to measure solar neutrinos, and later to resolve the solar neutrino
problem, began in the early days of nuclear astrophysics, with the first efforts
to understand proton burning in main sequence stars.
I would like to review that history, our current understanding of solar neutrinos, and
open questions in neutrino physics, and discuss some opportunities for further
solar neutrino measurements.

Solar neutrino physics brings together
stellar modeling, nuclear reactions, and observation.  A key early development
was the 1959 Holmgren and Johnston \cite{holmgren} measurement of the S-factor for the pp-chain reaction $^3$He($\alpha,\gamma)^7$Be, which proved to be surprisingly large.   This implied
that the sun could produce some of its energy through the more temperature dependent
ppII and ppIII cycles of the pp chain, elevating the neutrino fluxes expected 
from  $^7$Be electron capture
and $^8$B $\beta$ decay.  (See Fig. 1.)   These neutrinos contribute to ground- and
excited-state transitions in $^{37}$Cl($\nu$,e)$^{37}$Ar, a reaction for detecting 
neutrinos that had been proposed by Pontecorvo \cite{pontecorvo} in 1946 and 
considered by Alvarez \cite{alvarez} in 1949 (who studied backgrounds that
might inhibit, for example, a reactor neutrino experiment).  Alvarez's envisioned experiment -- proposed
as a test of the distinguishability of the neutrino
and antineutrino, prior to the discovery of parity violation  -- was later conducted by Davis at Savannah River 
\cite{jandr,davis55}.

\begin{figure}
\begin{center}
\includegraphics[width=12cm]{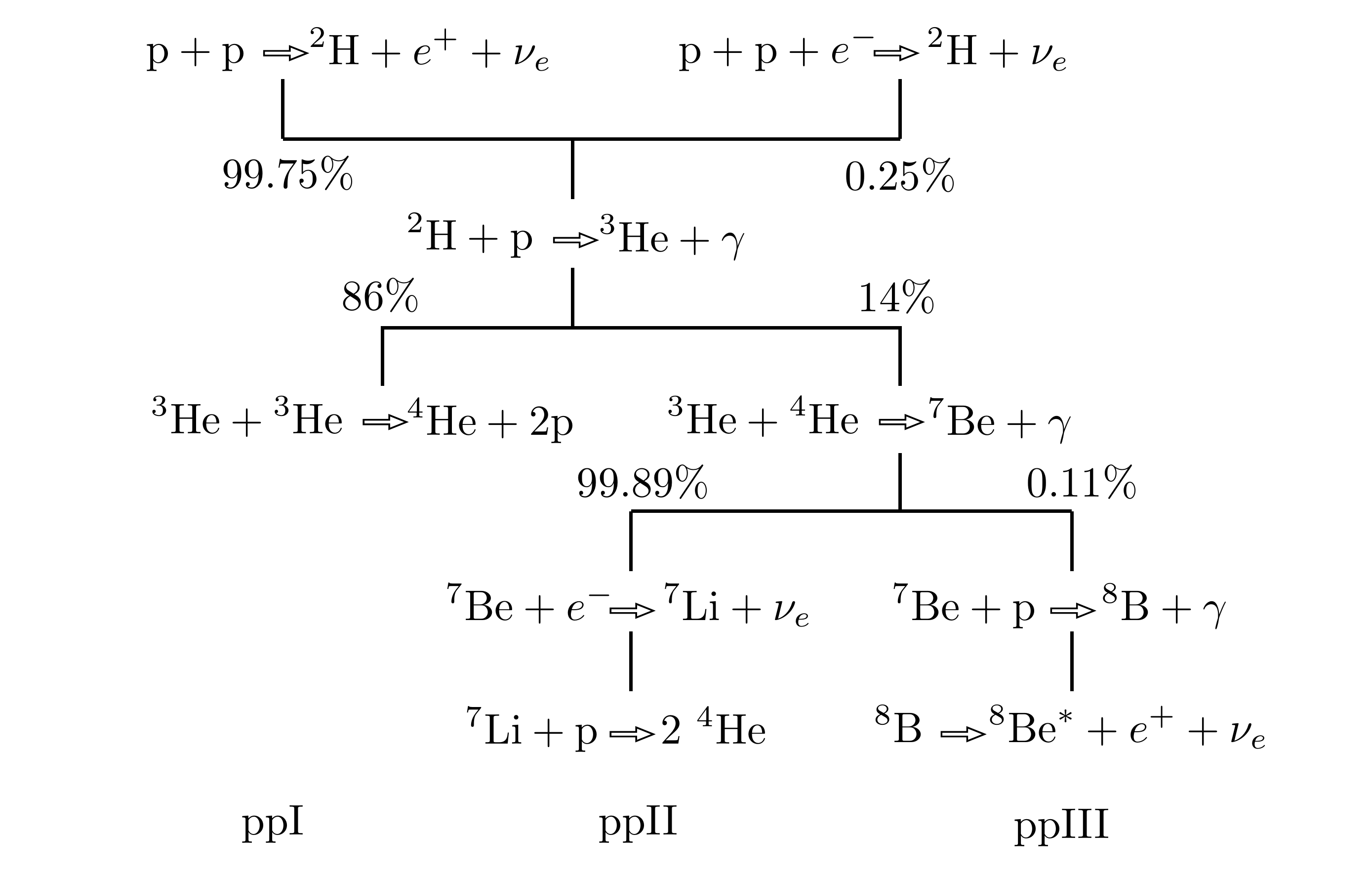}
\end{center}
\caption{The three cycles of the pp chain and associated neutrinos.}
\label{fig1}
\end{figure}

The Caltech effort in nuclear astrophysics brought three young researchers,
Bahcall, Iben, and Sears, together in the summer of 1962.   Stimulated in part
by the Holmgren and Johnston measurement, they began construction of a solar model to
predict the solar core temperature, the most important
parameter governing the competition between the ppI, ppII, and ppIII cycles, and to 
provide the first quantitative estimate of the
resulting neutrino fluxes.  The
Bahcall, Fowler, Iben, and Sears model, published in 1963, predicted a
counting rate for Davis's proposed 100,000-gallon chlorine solar neutrino
detector of about one event per day.

A key development occurred in 1963 when, during a seminar by Bahcall at Copenhagen,
Mottelson inquired about the importance of neutrino excitation of excited states
in $^{37}$Ar \cite{jandr}.   In 1964 Bahcall and Barnes \cite{barnes} pointed out that a calibration of the 
excited state contribution to the $^{37}$Cl
cross section could be made
by measuring the delayed protons from the analog $\beta$ decay of $^{37}$Ca,
the isospin mirror of $^{37}$Cl.  Effectively the lifetime of $^{37}$Ca would be a test of the
elevated cross section predicted on the basis of the excited-state contribution to $^8$B
neutrino absorption.  Subsequent measurements by Hardy and Verrall \cite{hardy} 
and Reeder, Poskanzer, and Esterlund \cite{poskanzer} 
established the importance of the excited-state contribution.
[These experiments were later repeated in a manner that was kinematically complete:
see Adelberger et al. \cite{adelberger} for a discussion.]

Bahcall \cite{bahcall64} and Davis \cite{davis64} published companion letters in March 1964 
arguing the adequacy and
feasibility of a 100,000-gallon Cl experiment to measure solar neutrinos.  Excavation of the
detector cavity in the Homestake Mine began in summer 1965.  The tank was installed
and filled by the next year.  The first results were announced by Davis, Harmer, and
Hoffman \cite{davis68} in 1968, an upper bound of 3 SNU (1 SNU = 10$^{-36}$ capures/Cl atom/sec), 
below the standard solar model (SSM) prediction of Bahcall, Bahcall, and Shaviv of
7.5 $\pm$ 3 SNU \cite{bbs}.

This result and associated theoretical work on suggested solutions led to a series of
experiments -- Gallex/GNO/SAGE \cite{gallex,gno,sage}, Kamiokande \cite{kamiokande}, 
Super-Kamiokande \cite{sk} , and the Sudbury
Neutrino Observatory \cite{sno}.  Efforts by Borexino \cite{borexino} and 
KamLAND \cite{kamland} to measure the $^7$Be
neutrinos are currently underway or in preparation.
These experiments are important as tests of the SSM and of the new neutrino physics
that proved to be the source of the solar neutrino problem.

\section{The Standard Solar Model}
The physics assumptions underlying the SSM include:
\begin{itemize}  
\item Hydrostatic equilibrium.  For each volume element it is assumed that
gravity is balanced by the gas pressure gradient.  This requires specification of the 
electron gas equation of state as a function of temperature, heavy element abundance
Z, and density.  The EOS is very nearly that of an ideal gas.
\item Energy transport.  The sun has a radiative interior and convective
envelope, with the location of the boundary sensitive to the modeling of the radiative
opacity.  The depth of the convective zone can be determined experimentally,
because it influences solar surface acoustic modes (helioseismology).
\item Energy generation.  Solar energy is generated by the conversion of four
protons to $^4$He with the release of about 25 MeV in energy, with the pp-chain accounting
for nearly 99\% of the reactions (and the CNO cycle the remainder).   The main laboratory task has been determining the nuclear cross sections for the various reactions
to sufficient precision.  Because typical center-of-mass energies in the solar core are $\sim$ 2 keV,
in general this requires measuring reactions at higher energies, then using 
r-matrix or other models to extrapolate the laboratory S-factors to threshold.
\item Boundary conditions.  The sun's age and current luminosity, radius, mass, and
surface composition are known.   While the composition of the solar core at the onset
of the main sequence is not known directly, the SSM assumes that the zero-age core
metallicity $Z$ can be equated to today's surface value.  As the mass fractions in H,
He, and heavy elements Z must sum to one, a single additional constraint is needed.
This is the solar luminosity: the zero-age H/He ratio is adjusted until the correct luminosity
is achieved after 4.6 b.y. of evolution.
\end{itemize}

The nuclear physics efforts on the pp-chain have reach a very high level of sophistication \cite{parker}.
Long the most uncertain rate in the pp-chain, the S-factor for $^7$Be(p,$\gamma$) has
now been determined to an accuracy of $\lsim$ 5\%
\begin{equation}
S_{17}(20~\mathrm{keV}) = 20.6 \pm 0.5 \pm 0.6 ~\mathrm{eV-b} 
\end{equation}
by a series of six ``direct" measurements \cite{junghans}.  Also
notable are the measurements by the LUNA collaboration, working in the low-background
environment of Gran Sasso, on $^3$He+$^3$He.  The group succeeded in obtaining data in
the solar Gamow peak, thus largely eliminating theoretical uncertainties in the extrapolation
of high-energy data to solar energies \cite{bonetti}.

The SSM has evolved over the years.  For example, the growing accuracy of helioseismology
helped motivate efforts to include the diffusion of He and heavy elements over the solar
lifetime.  Important checks on the SSM include the predicted depth of the convective zone,
which is constrained experimentally by the frequency distribution of low-$l$ acoustic
modes.  The model that emerges is dynamic: there is a $\sim$ 44\% luminosity growth
over the solar lifetime due to changing core chemistry and thus opacity.  The high-energy
$^8$B neutrino flux is a relatively contemporary phenomenon: it is predicted to grow
with a time constant $\tau_0 \sim 0.9$ b.y., $\phi(^8\mathrm{B}) \sim \phi_0 e^{-t/\tau_0}$.
Yet the model remains somewhat limited in scope.  As calculations are done in 1D, there is
no attempt to model the detailed behavior of the convective zone or of the convective onset
of main-sequence burning.  Known phenomena such as the depletion of surface Li are
presumably connected with such physics.

Solar neutrinos were initially viewed -- and remain -- an important test of the SSM.  The
three cycles making up the pp-chain (see Fig. 1) are tagged by neutrinos -- the pp neutrino
flux constrains the overall rate of H burning, while the $^7$Be and $^8$B neutrino
fluxes can be used to determine the ppII and ppIII rates, respectively.  Because of Coulomb barriers, the competition
between the three cycles depends rather sensitively on the solar core temperature.  Thus
solar neutrino flux measurements -- given the quality of current 
calibrations of the nuclear microphysics  -- can
fix the core temperature to an accuracy of about 1\%.

But, by the mid-90s, instead of such a temperature determination, an important discrepancy
had been confirmed.  The combination of the Cl, gallium, and Kamioka experiments 
seemed to require (assuming otherwise standard physics)
\begin{equation}
\phi^{exp}(\mathrm{pp}) \sim \phi^{SSM}(\mathrm{pp})~~~\phi^{exp}(^7\mathrm{Be}) \sim 0~~~\phi^{exp}(^8\mathrm{B}) \sim 0.4 \phi^{SSM}(^8\mathrm{B})
\end{equation}
The difficulty posed by these results can be easily seen.  A low $\phi^{exp}(^8 \mathrm{B})$ naively
implies a solar core cooler than predicted by the SSM: due to the $\sim T^{24}_C$ 
dependence of the $^8$B neutrinos, a core temperature $\sim 0.96 T_C$ would account for
the Kamioka and Cl results.  However the SSM also predicts that $\phi(^7\mathrm{Be} )/\phi(^8\mathrm{B}) \sim T_C^{-14}$.  Thus the low value of this ratio, compared to SSM predictions,
requires a hotter core.   While such an analysis based on a single parameter $T_C$ may
seem naive, more detail investigations came to the similar conclusions.  For example, Fig. 2, from
Castellani et al. \cite{cast}, shows that a variety of SSM perturbations yield fluxes
that corresponding to expectations based on $T_C$.

\begin{figure}
\begin{center}
\includegraphics[width=11cm]{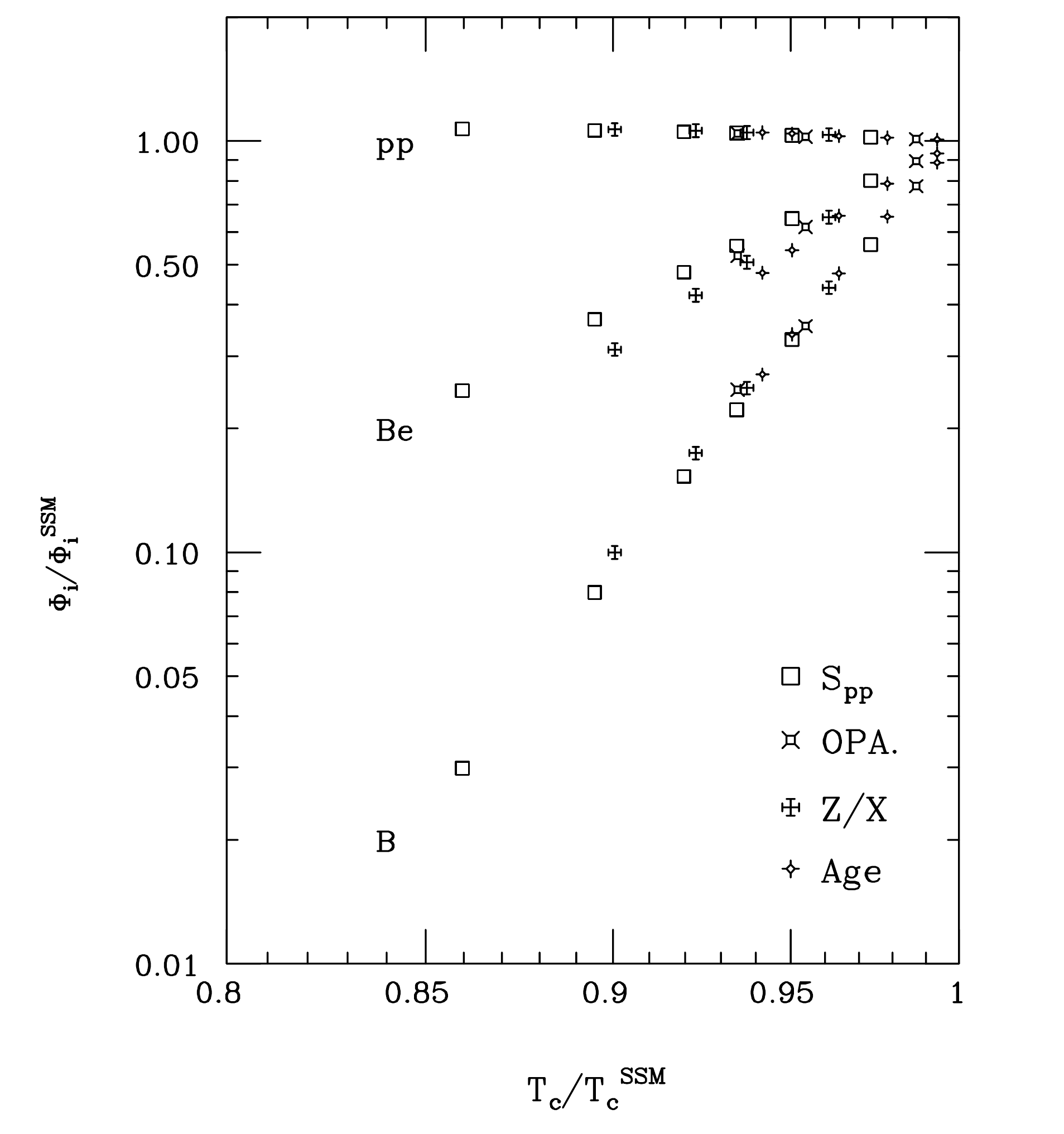}
\end{center}
\caption{Neutrino fluxes resulting from changes in various SSM parameters, such as
the p+p S-factor, opacity, heavy element abundance, and solar age.  The fluxes 
correlate very well with one parameter, the resulting core temperature $T_C$.
This naive $T_C$ dependence was recognized early on to be incompatible
with the pattern of fluxes deduced from the Cl, Gallex/SAGE, and Kamioka
experiments.  From Castellani et al. \cite{cast}.}
\label{fig2}
\end{figure}

\section{Massive Neutrinos}
The standard model (SM) of particle physics has massless neutrinos.  But if extended -- treated
as an effective theory with a dimension-full Majorana mass term (the only dimension-five operator
that can be constructed with SM fields) or enlarged to include the right-handed
neutrino field needed for a Dirac mass term -- neutrinos would be massive.  Massive neutrinos
not coincident with their flavor eigenstate counterparts will lead to the phenomenon of
neutrino oscillations, as Pontecorvo first pointed out.  Furthermore, fascinating new oscillation
phenomenon can occur because solar neutrinos are created at high density in the solar
core, then propagate to low density.  Mikheyev and Smirnov \cite{ms} showed that the effective mass
neutrinos acquire in matter, a phenomenon first discussed by Wolfenstein \cite{lincoln}, could lead to 
large oscillation probabilities, even for small vacuum mixing angles.   Such matter-enhanced
neutrino oscillations can be viewed as a level-crossing phenomenon:  the local masses
of the neutrinos reflect the surrounding electron density, with the electron neutrino becoming
heavier at high density.  If the solar core density is sufficient to cause a level inversion, then
a level crossing will occur somewhere in the sun, as the neutrino propagates outward \cite{bethe}.
If that crossing is adiabatic \cite{haxton,parke}, then strong $\nu_e \rightarrow \nu_\mu$
conversion will occur -- as illustrated in Fig. 3.  Details of  this process are discussed in many places, and will not be repeated here.

\begin{figure}
\begin{center}
\includegraphics[width=12cm]{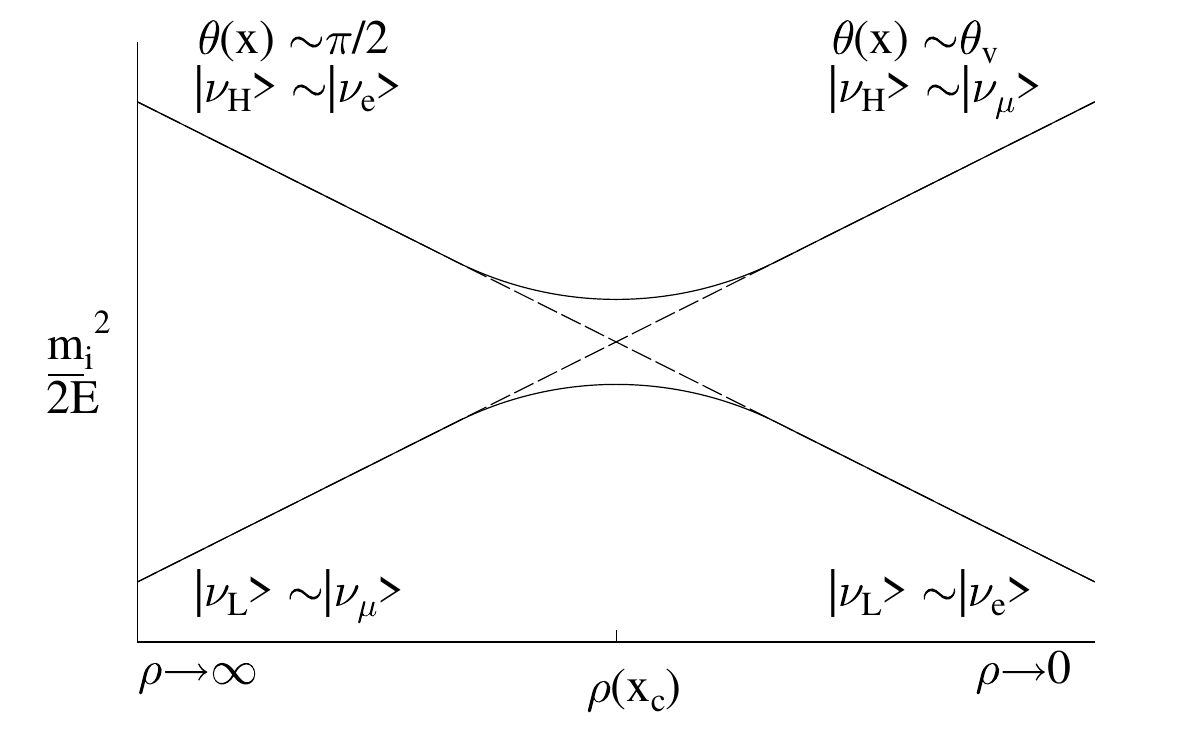}
\end{center}
\caption{Schematic illustration of the MSW avoided level crossing that may arise because
the $\nu_e$'s effective mass increases in matter.  In the adiabatic limit, a neutrino would
follow one of the local mass eigenstate trajectories depicted by the solid lines.  A $\nu_e$ 
created in the solar core as the heavy mass
eigenstate can leave the sun as the vacuum heavy eigenstate which, if the vacuum
mixing angle is small, may be nearly coincident with the $\nu_\mu$.}
\label{fig3}
\end{figure}

The early solar neutrino results and the possibility of discovering new particle physics 
provided the impetus for two important direct-counting experiments, Super-Kamiokande \cite{sk} and the
Sudbury Neutrino Observatory (SNO) \cite{sno}.  SNO, a Cerenkov detector with an inner vessel
containing a kiloton of heavy water surrounded by seven kilotons of light water, was
unique in its sensitivity to neutrino flavors.  It and Super-Kamiokande both made use of 
$\nu_x-e$ elastic scattering (ES)
\begin{equation}
\nu_x + e^- \rightarrow \nu_x^\prime + e^{-}, 
\end{equation}
a reaction that takes place for both electron and heavy-flavor neutrinos, with cross sections
roughly in a 6:1 ratio.   The scattered electrons are forward peaked, allowing the experimenters
to cut away backgrounds by correlating signals with the position of the sun.
But SNO also detected two other reactions, the charged-current (CC) and neutral-current (NC)
breakup of deuterium,
\begin{eqnarray}
\nu_e + d \rightarrow p + p + e^- \nonumber \\
\nu_x + d \rightarrow \nu_x^\prime + n + p ~.
\end{eqnarray}
These reactions are sensitive, respectively, just to $\nu_e$s or equally to neutrinos of any flavor.  The
CC reaction was detected via the scattered electron, which tends to carry off most of the
neutrino energy (helpful in reconstructing the neutrino spectrum) but is emitted almost
isotropically (so there is little directionality that can be exploited to reduce backgrounds).
The signal for the NC reaction is the neutron, which was observed in SNO by $(n,\gamma)$
capture on a Cl-bearing salt added to the detector and, later, in tubular $^3$He
proportional counters that were installed in the detector.   SNO's
great depth ($\sim$ 6 km water equivalent) and clean-room operating standards
made such measurements possible,
reducing cosmic and environmental radioactivity backgrounds to very low levels.

\begin{figure}
\begin{center}
\includegraphics[width=12cm]{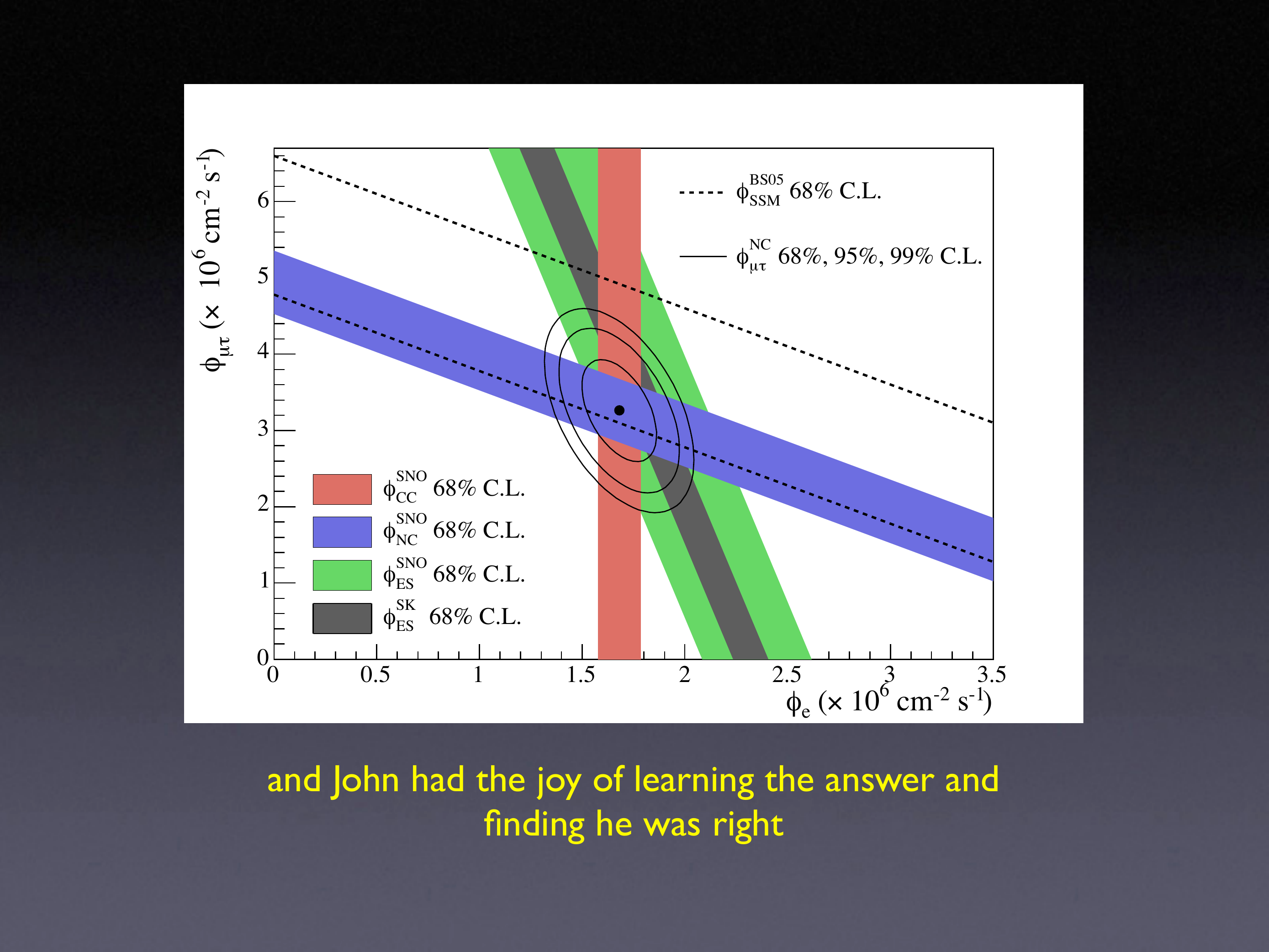}
\end{center}
\caption{The SNO diagram \cite{sno} showing the region consistent with the ES, CC, and NC 
measurements.  Also shown are the Super-Kamiokande ES results \cite{sk} and (designated
by dashed lines) the band corresponding to the SSM prediction for the total $^8$B
neutrino flux.}
\label{fig4}
\end{figure}

The SNO results are shown in Fig. 4.  The bands representing the three neutrino detection
channels, with their very different sensitivities
to neutrino flavor, converge at an ellipse indicating that approximately two-thirds of the
solar neutrinos arrive on earth as heavy-flavor neutrinos.  Thus the discrepancy first noted
by Davis, whose Cl detector recorded $\nu_e$s only, was not due to an incorrect
estimate of the solar neutrino flux, but rather to their partial conversion to other flavors
during transit to earth.  Indeed, Fig. 4 shows that the SSM prediction for the $^8$B solar
neutrino flux is in good agreement with the flavor-blind NC measurement made in SNO.

\section{Next Steps?}
Thus the question, where does the solar neutrino field go from here?  Three important
directions are:
\begin{itemize}
\item Pursuit of several important open questions about neutrino properties, using both
accelerator/reactor and astrophysical neutrino sources.  These questions are important to
the modeling of a variety of exotic stellar environments, such as core-collapse supernovae, 
and to the construction of extensions to the SM that will encompass the
new neutrino physics.
\item Completing the spectroscopy of pp-chain neutrinos:  This includes direct measurements,
such as those underway by Borexino \cite{borexino} and KamLAND \cite{kamland}, 
of the $^7$Be neutrinos, and future
experiments to determine the flux and flavor of the dominant solar neutrino source, the
pp neutrinos.  Existing constraints on the low-energy fluxes come from the radiochemical
Cl and Gallex/GNO/SAGE detectors. 
\item Measuring the CNO neutrinos.  I will argue below that such a measurement is not
only important to understanding the sun -- a CNO neutrino measurement would determine
directly the solar core metallicity -- but also to the general theory of main-sequence 
evolution of massive stars.
\end{itemize}

\begin{figure}
\begin{center}
\includegraphics[width=12cm]{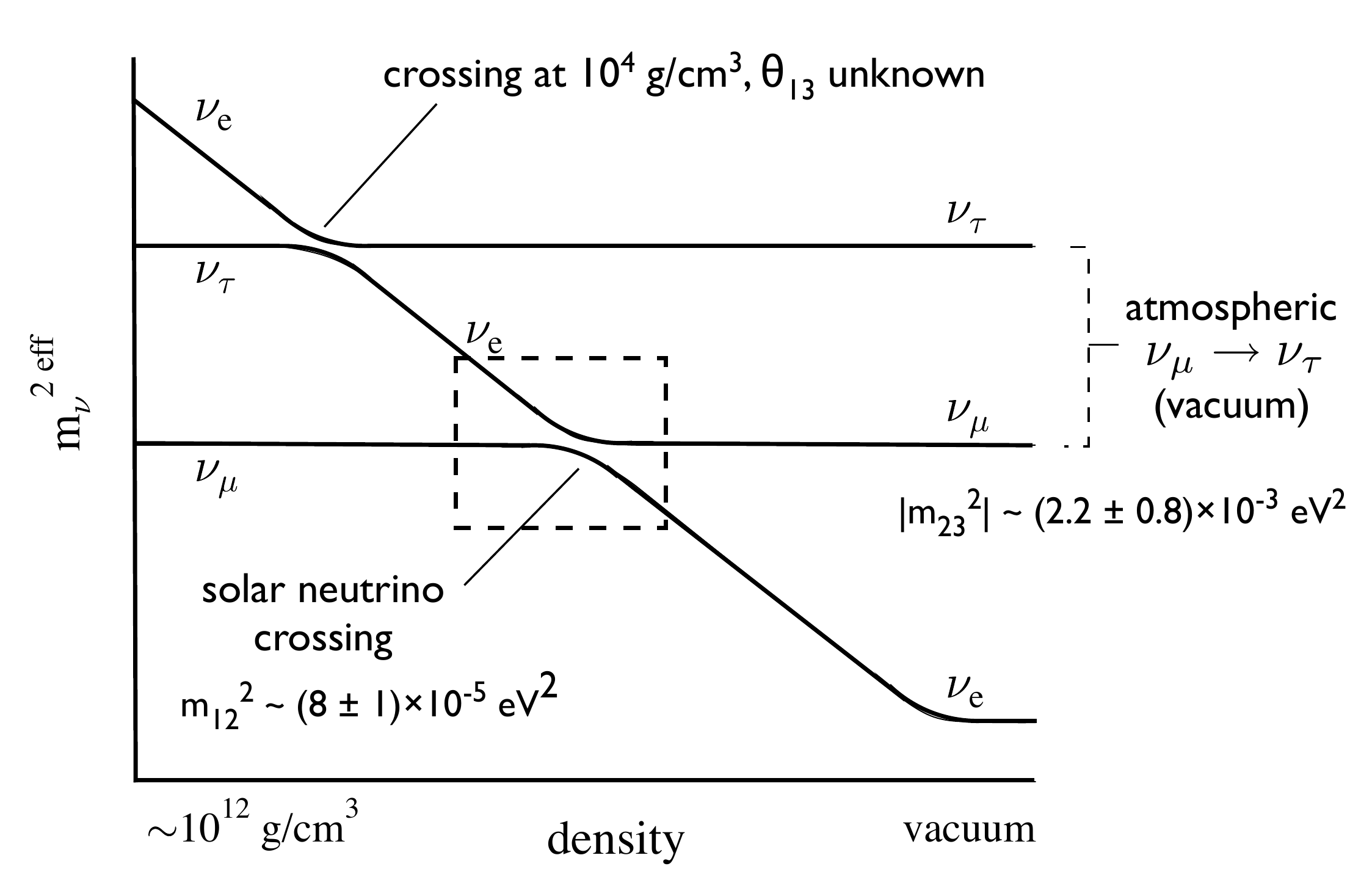}
\end{center}
\caption{A three-flavor plot showing two MSW crossings, the second of which corresponds
to a density of $\sim 10^4$ g/cm$^3$, typical of the carbon zone of the progenitor of a 
Type II supernova.  The relevant mixing angle $\theta_{13}$ has not yet been measured.}
\label{fig5}
\end{figure}

\noindent
{\it Neutrino properties:} Figure 5 illustrates some of the expectations for neutrino physics
in stellar environments, assuming a regular (rather than inverted) hierarchy where the
two neutrino states participating in solar neutrino mixing are lighter than the third state.  The
physics discussed in connection with solar neutrinos -- the 1-2 level crossing arising
from matter effects --
is repeated at higher density, in a second crossing.  The atmospheric $\delta m_{23}^2 \sim
(2.2 \pm 0.8) \times 10^{-3}$ eV$^2$ leads to the expectation that for neutrino energies
typical of a supernova ($\sim$ 10 MeV),  this crossing would be encountered in the
carbon zone of a massive star, where $\rho_e \sim 10^4$ g/cm$^3$.  However, the
relevant mixing angle $\theta_{13}$ is not known: the current upper bound from the
Chooz reactor experiment is $\sin^2{2 \theta} \lsim 0.19$ (at $\delta m^2$ = 2.0 $\times 10^{-3}$
eV) \cite{chooz}.  This crossing
(as well as additional neutrino background effects discussed by George Fuller at this meeting)
has the potential to alter energy deposition in a supernova, by causing the exchange
of cooler electron neutrinos and hotter heavy-flavor neutrinos.  Hotter $\nu_e$s 
increase the neutrino-matter coupling.  Such an inversion could
produce distinctive signatures in terrestrial detectors with sensitivities to different
neutrino flavors.
As the crossing is expected to remain adiabatic for mixing angles $\gsim 10^{-4}$,
this signature could prove very important if $\theta_{13} \lsim 0.01$, the level that 
proposed experiments such as Double Chooz \cite{doublechooz} and Daya Bay \cite{daya}
are expected to reach.  Prior to construction of a neutrino factory, terrestrial experiments
may not be able to reach much beyond the 0.01 level.

\begin{figure}
\begin{center}
\includegraphics[width=12cm]{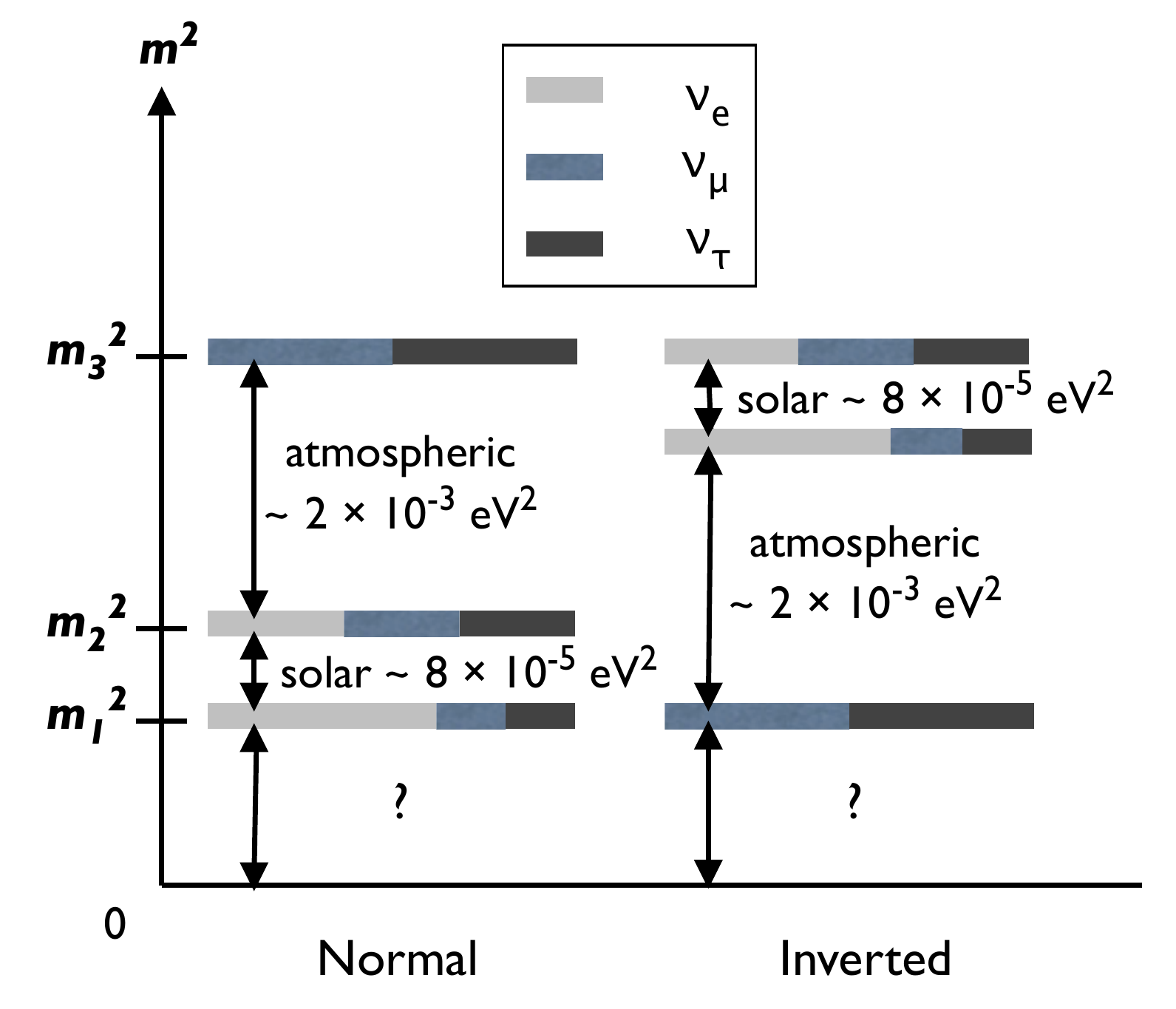}
\end{center}
\caption{The flavor content, allowable hierarchies, and unknown mass scale of the
neutrinos.}
\label{fig6}
\end{figure}

Figure 6 illustrates some of remaining questions related to masses.  We know the solar neutrino
mass-squared difference $\delta m_{12}^2=m_2^2-m_1^2$.  In the case of
atmospheric neutrinos, the magnitude, but not the sign, of 
$\delta m_{23}^2$ is known.  Thus two hierarchies -- normal and inverted -- are allowed by the data.
The absolute scale of neutrino masses -- the offset from zero shown in Fig. 6 -- is limited 
by laboratory experiments (tritium $\beta$ decay) to $\lsim 2.2$ eV \cite{mainz}.  The behavior of the
neutrino mass under particle-antiparticle conjugation is not known: extensions of the
SM would allow both Dirac and lepton-number-violating Majorana terms, 
and indeed the presence of both is exploited in the seesaw mechanism to explain the
smallness of neutrino mass relative to other SM fermions,
\begin{equation}
m_\nu \sim m_D \left[ {m_D \over M_R} \right], 
\end{equation}
where $m_D$ is a typical SM Dirac mass and $M_R$ a heavy right-handed neutrino mass.  That
is, $m_\nu$ is suppressed relative to other SM masses by the small parameter $m_D/M_R$.
Indeed, an $M_R \sim 0.3 \times 10^{15}$, near the GUT scale, is suggested by the identification
of $m_3 \sim \sqrt{\delta m_{23}^2} \sim 0.05$ eV.  

Progress will be made on each of these questions in the next decade.  Next-generation
long-baseline neutrino experiments have, as one of their goals, the use of
matter effects to distinguish between the normal and inverted hierarchies \cite{baseline}.
The new tritium $\beta$ decay experiment KATRIN has the potential to tighten
the limit on $m_{\nu_e}$, and thus the absolute scale of neutrino mass, by almost an order of
magnitude \cite{katrin}. 
Similarly, cosmological analyses that calculate the effects of
massive neutrinos on large-scale structure, as deduced from galaxy surveys and from temperature
fluctuations in the cosmic microwave background (CMB), place a constraint on the neutrino mass
contribution to the closure density.  Currently this yields
\begin{equation}
\sum_i m_\nu(i) \lsim 0.7 ~\mathrm{eV}, 
\end{equation}
though this limit is expected to tighten significantly when Planck and other future CMB observatories produce data.

Several next-generation neutrinoless double $\beta$ decay experiments are
under development, with the ultimate goal of reaching Majorana mass sensitives of
\begin{equation}
 \left| \sum_{i=1}^{2n} \lambda_i U_{ei}^2 m_i \right| \lsim 0.05 ~\mathrm{eV},
 \label{bb}
 \end{equation}
the scale we noted was set by $\sqrt{\delta m_{23}^2}$.  Here the sum extends over neutrino
generations, with each neutrino mass eigenstate $i$ contributing in proportion to its mass and
to its coupling
probability to the electron $U_{ei}^2$.  Each term is weighted by a phase $\lambda_i$
that, if CP is conserved, corresponds to $\pm i$: thus mass eigenstates will tend to cancel in this sum
if they have opposite CP.  If CP is violated, the $\lambda_i$ also include the effects of two 
Majorana phases that will be difficult to constrain by other means.  Double beta decay is
our most powerful probe of lepton number violation and Majorana masses, and has 
potentially the most reach as a laboratory test of neutrino mass.  It can
distinguish among competing neutrino mass scenarios, e.g., quasi-degenerate schemes
vs. schemes where the electron neutrino is quite light, though the mass defined by Eq. (\ref{bb})
is not simply related to the kinematic mass measured in experiments like tritium $\beta$ decay.

Denoting the mass eigenstates by $\nu_1,~\nu_2,$ and $\nu_3$, the relationship between the mass and flavor eigenstates is a product of rotations in the 2-3 (atmospheric neutrino), 1-3 (reactor $\bar{\nu}_e$
disappearance), and 1-2 (solar neutrino) subspaces.   Defining the respective angles by
$\theta_{23}$, $\theta_{13}$, and $\theta_{12}$, with $\cos{\theta_x} \equiv c_{x}$ and
$\sin{\theta_{x}} \equiv s_{x}$, the mixing matrix is
\begin{eqnarray}
\left( \begin{array}{c} \nu_e \\ \nu_\mu \\ \nu_\tau \end{array} \right)
&=& \left( \begin{array}{ccc} 1 & & \\ & c_{23} & s_{23} \\ & -s_{23} & c_{23} \end{array} \right)
\left( \begin{array}{ccc} c_{13} & & s_{13} e^{-i \delta} \\ & 1 & \\ -s_{13} e^{i \delta} & & c_{13} \end{array} \right)
\left( \begin{array}{ccc} c_{12} & s_{12} & \\ -s_{12} & c_{12} & \\ & & 1 \end{array} \right)
\left( \begin{array}{c} \nu_1 \\ e^{i \phi_2}\nu_2 \\ e^{i \phi_3}\nu_3 \end{array} \right) \nonumber \\
&& \nonumber \\
&=& \left( \begin{array}{ccc} c_{12}c_{13} & s_{12}c_{13} & s_{13} e^{-i \delta} \\
-s_{12}c_{23}-c_{12}s_{23}s_{13} e^{i \delta} & c_{12}c_{23}-s_{12}s_{23}s_{13} e^{i \delta} & s_{23}c_{13} \\
s_{12}s_{23}-c_{12}c_{23}s_{13} e^{i \delta} & -c_{12}s_{23}-s_{12}c_{23}s_{13} e^{i \delta} & c_{23}c_{13} 
\end{array} \right) 
\left( \begin{array}{c} \nu_1 \\ e^{i \phi_2}\nu_2 \\ e^{i \phi_3}\nu_3 \end{array} \right)~.
\end{eqnarray}
This matrix includes three CP-violating phases, the Dirac phase $\delta$ and two Majorana
phases $\phi_2$ and $\phi_3$.

The atmospheric and solar neutrino experiments have determined $\theta_{23} \sim 45^\circ$
and $\theta_{12} \sim 30^\circ$ and, while greater accuracy is always important (especially to
determine how close $\theta_{23}$ might be to the maximal mixing limit of $\pi/4$), most
attention is now focused on determining the unknown parameters in the mixing matrix.
Perhaps most important is the third mixing angle $\theta_{13}$, which, as noted earlier,
is so far only bounded by reactor neutrino results.  New reactor experiments currently in
preparation, Double Chooz and Daya Bay, are designed to reach sensitivities of approximately
0.02 and 0.008, respectively.

Given that the source of CP violation responsible for the excess of matter over antimatter
in our universe is still uncertain, the determination of the scale of leptonic CP violation is also
a major goal.  With suitable attention to matter effects associated with neutrino beams passing
through the earth and to other parameter degeneracies \cite{barger}, CP violation can be
determined in long-baseline neutrino oscillation experiments that compare $P[\nu_\mu \rightarrow
\nu_\tau]$ with $P[\bar{\nu}_\mu \rightarrow \bar{\nu}_\tau]$.  The CP violation
is proportional to the Jarlskog invariant
\begin{equation}
\sin{2 \theta_{12}} \sin{2 \theta_{23}} \sin{2 \theta_{13}} \cos{\theta_{13}} \sin{\delta} .
\end{equation}
The first two factors are
known to be large.  Thus a demonstration that $\theta_{13}$ is not too small would imply
significant sensitivity to CP violation.  Probing such CP violation is one of the major
goals for very long baseline neutrino oscillation experiments, where the relative size of the CP-violating
observable grows with distance \cite{marciano}.\\

\noindent
{\it Low-energy neutrino spectroscopy:}  Pending results from Borexino and KamLAND on their
efforts to measure $^7$Be neutrinos, direct real-time measurements are so far limited to
the $^8$B $\nu$ spectrum above 5 MeV.  Thus 99.99\% of the flux has not been detected by
direct means, including the pp/pep and $^7$Be neutrinos that tag the ppI and ppII cycles
(see Fig. 7, taken from Bahcall \cite{johnfig}).

\begin{figure}
\begin{center}
\includegraphics[width=12cm]{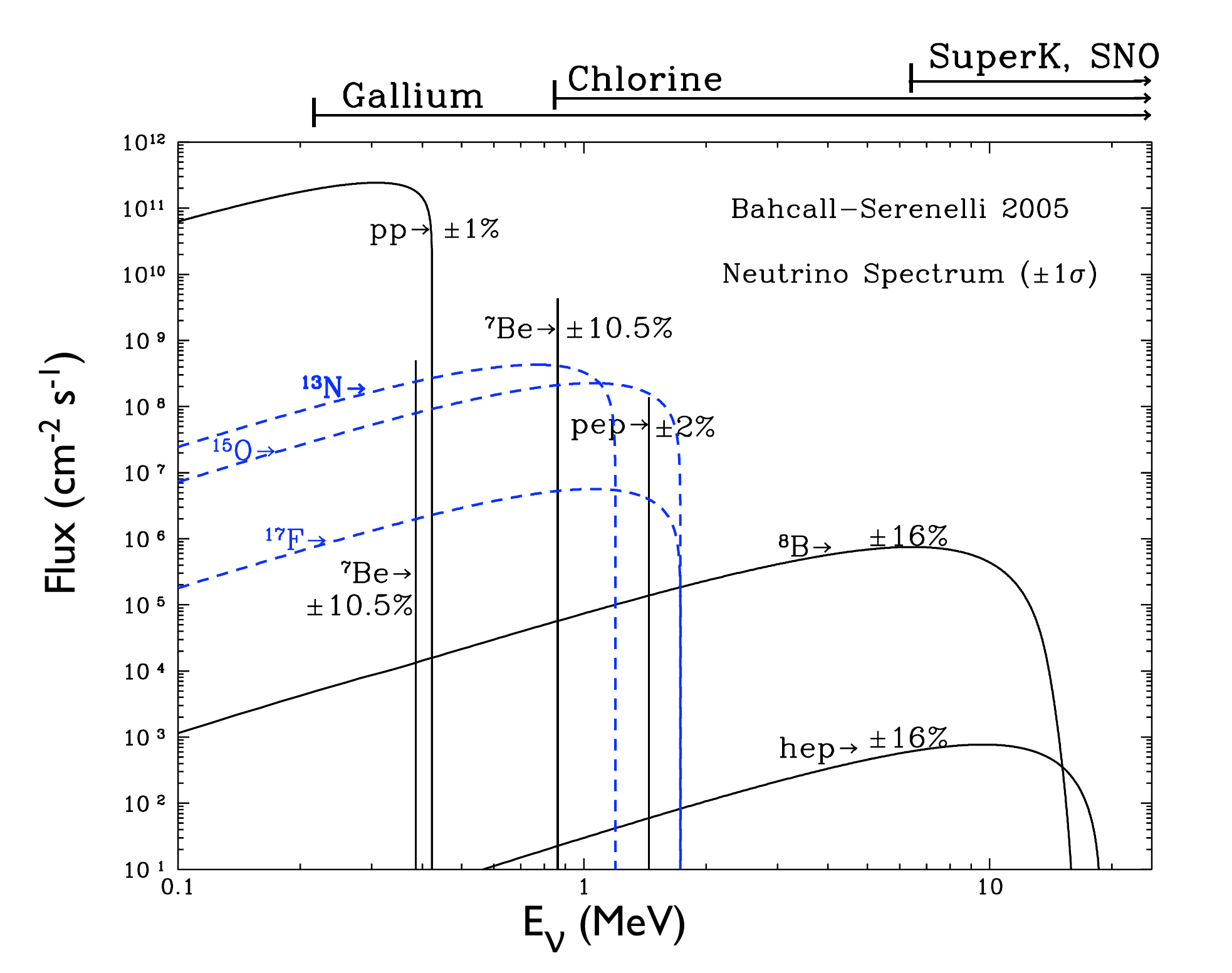}
\end{center}
\caption{The spectrum of solar neutrinos, from Bahcall \cite{johnfig}.  SNO and Super-Kamiokande
spectral measurements were limited to neutrinos with energies above 5 MeV.  Borexino will
make the first direct measurement of sub-MeV neutrinos.  The CNO fluxes could be identified
in an experiment like SNO+.}
\label{fig7}
\end{figure}

The oscillation parameters deduced from global fits to solar neutrino data indicate that 
the solar $\nu_e$ spectrum
will be distorted by matter effects.  Yet so far there has been no
direct measurement of the energy dependence in the survival
probability $P(E_\nu)$ or of MSW day-night effects due to neutrino passage through the earth.
The oscillation parameters indicate that
the level-crossing boundary of the 1-2 MSW triangle will be encountered at a neutrino energy
of $\sim$ 3 MeV.  Thus one can explore the transition from vacuum oscillations
to matter oscillations by mapping $P(E_\nu)$ from low to high neutrino energies.  Borexino, now operating, will determine $P(E_\nu=0.86~\mathrm{MeV})$,
for example.

A variety of CC and NC pp-neutrino detectors are under development.
In addition to the exploration of matter effects, these detectors will be able to exploit 
nature's most intense and well-characterized source of $\nu_e$s: the solar pp flux and spectrum
are known to an accuracy of about 1\%.  While experiments like KamLAND have succeeded
in reducing uncertainties on $\delta m_{12}^2$ by measuring $P(\bar{\nu}_e)$ at different
baselines, uncertainties in the reactor $\bar{\nu_e}$ spectrum limit the accuracy of 
$\theta_{12}$ determinations.  But solar pp-neutrino measurements could, in principle,
determine this angle to $\sim$ 1\%. \\

\noindent
{\it Testing Stellar Modeling: The CNO Neutrinos:}  Despite the minor role CNO neutrinos
play in our sun, there is strong motivation for exploiting these neutrinos as a quantitative
test of our understanding of the CNO cycle:
\begin{itemize}
\item The CNO cycle, due to its sharper dependence on stellar core temperature, is the reaction chain
that sustains massive main-sequence stellar evolution.  This dependence is reflected in
the corresponding neutrino fluxes, which vary as $\phi(\mathrm{CNO}) \sim T_C^{24-27}$.  
Because SNO and Super-Kamiokande
have made accurate measurements of the most temperature-dependent component of
the pp-chain ($\phi(^8\mathrm{B}) \sim T_C^{24}$), we know the core temperature in our sun to an accuracy of about 1\%.
Thus we can now use the solar core as a calibrated laboratory in which to test our
understanding of the CNO cycle.
\item The CNO cycle is important to other systems of current interest, such as the
first massive metal-poor stars (where hydrogen burning via the CNO cycle
turns on only after an early phase
of $3\alpha \rightarrow ^{12}$C produces metals).
\item One of the principal assumptions of the SSM is the identification of today's
surface metal abundances with the zero-age core metallicity.  A measurement of CNO
neutrinos is a direct check on this assumption, as the flux is proportional to core
metallicity.
\item This issue -- using surface observations to constrain the core -- is now central to
the principal anomaly in the SSM, that new surface metal determinations have placed
helioseismology results (e.g., deduced sound speeds, estimates of the depth of the convective zone) in conflict with SSM predictions \cite{asplund}.
\item The CNO cycle is thought to be responsible for an early convective stage in our sun,
extending about 10$^8$ years, driven by out-of-equilibrium burning.  Current efforts to
build 2D/3D extensions of the SSM are an important step toward modeling the early sun.
Thus one would like to verify that equilibrium CNO-burning is understood, to establish 
a foundation for later studies of solar convection and out-of-equilibrium burning.
\item The nuclear physics of the CNO cycle has been put on much firmer ground due to 
recent measurements of the controlling cross section.

\end{itemize}
\begin{figure}
\begin{center}
\includegraphics[width=12cm]{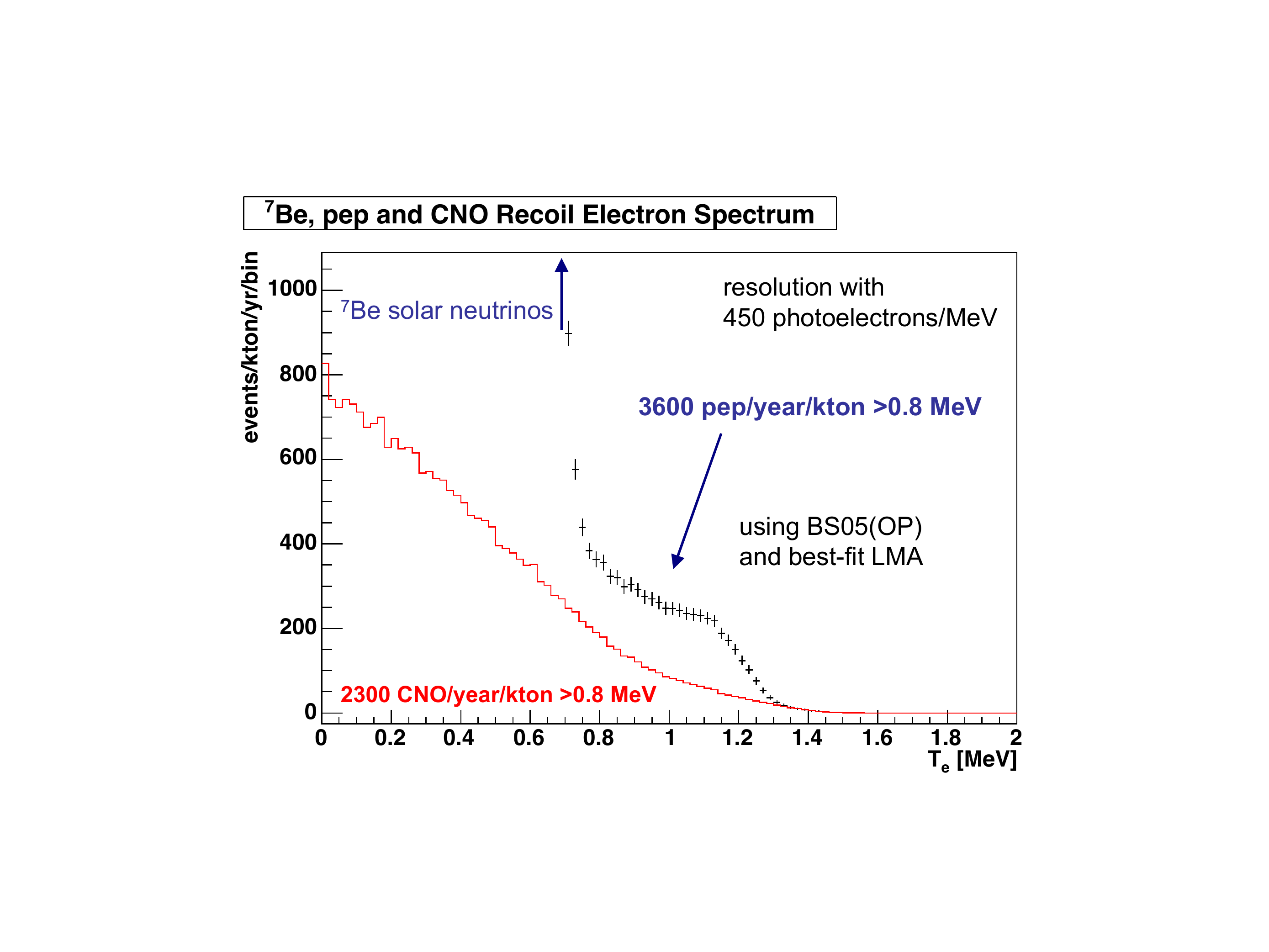}
\end{center}
\caption{The recoil spectrum expected from elastic scattering of pep, $^7$Be, and CNO
neutrinos in SNO+.  Figure from M. Chen \cite{chen}.}
\label{fig8}
\end{figure}

The sharp $T_C$-dependence of CNO hydrogen burning is
important to the stability of massive stars.  While a minor contributor to SSM energy 
production ($\sim$ 1\%), the CNO cycle does produce measurable solar neutrino fluxes \cite{bp04}
\begin{eqnarray}
{}^{13}\mathrm{N} &\rightarrow& {}^{13}\mathrm{C} + e^+ + \nu_e~~~~E_{\mathrm{max}} = 1.199 \mathrm{MeV}~~~~
\phi \sim 5.7 \times 10^8/\mathrm{cm}^2\mathrm{s} \nonumber \\
{}^{15}\mathrm{O} &\rightarrow& {}^{15}\mathrm{N} + e^+ + \nu_e~~~~E_{\mathrm{max}} = 1.732 \mathrm{MeV}~~~~
\phi \sim 5.0 \times 10^8/\mathrm{cm}^2\mathrm{s} .
\end{eqnarray}
To predict the response of terrestrial detectors to these sources one needs $T_C$ (calibrated
in the $^8$B flux measurements
by SNO and Super-Kamiokande), the oscillation parameters $\delta m_{12}^2$ and $\theta_{12}$,
and the nuclear cross sections for the CNO cycle.
Recent progress has also been made on nuclear physics: new
measurements by the LUNA collaboration \cite{luna} and at TUNL \cite{tunl} have
reduced uncertainties in the rate-controlling ${}^{14}$N(p,$\gamma$) cross section.  LUNA 
measured the S-factor down to 70 keV, finding a result that is 50\% smaller than the
previous ``best value."  This revision has had a significant impact on stellar age
determinations, pushing back
globular cluster ages by an estimated 0.7-1.0 b.y.

There is a new idea for building a 
high-counting-rate detector sensitive to CNO neutrinos,
construction of a large-volume scintillation detector in the cavity previously
occupied by SNO.  The new detector, SNO+ \cite{chen}, if developed for solar neutrinos
(in addition to double beta decay), would be able to detect about 2300 CNO neutrinos/year/kton,
above a threshold of about 0.8 MeV, as shown in Fig. 8.  It would appear that a flux measurement
accurate to $\sim$ 10\% might be possible.

The combination of a practical experiment, more certain nuclear physics, a calibrated
solar core temperature, and known 
neutrino parameters $\delta m_{12}$ and $\theta_{12}$ appear to make a direct
measurement of core metallicity possible.  The new surface abundances that have been
derived from improved 3D atmospheric solar absorption line analyses are difficult to
dismiss, despite the tension they have generated between the SSM (and its neutrino
predictions) and helioseismology.  The new abundances generally bring the sun
into better accord with galactic composition trends \cite{turck-chieze}.  It may turn
out that the SSM assumption that equates surface and zero-age core metallicity is
unjustified -- that some evolutionary effect breaks this equivalence.  In any case, the
opportunity to test this assumption experimentally should be taken.

\section{Summary}
Neutrino astrophysics and the theories of the origin of the elements, the main theme
of this conference, share a common history.  Laboratory astrophysics has made solar neutrino
physics into a quantitative field, and allowed experimenters to anticipate the kinds of major
discoveries that justified experiments like SNO and Super-Kamiokande.  The results
 -- discovery of neutrino mass and flavor mixing characterized by large angles -- are of
great importance, providing our first constraints on physics beyond the SM
of particle physics.  But as summarized here, the list of remaining laboratory 
neutrino physics questions is long.  The answers to the open questions will be important in
helping us characterize extreme astrophysical and cosmological neutrino environments.  
The needed 20-year program of laboratory and astrophysical neutrino studies is not unlike the
laboratory/astrophysics interface that Willie Fowler cultivated to help us understand the
origin of the elements.

Despite the current focus on particle physics properties of neutrinos, solar neutrino
spectroscopy remains an important probe of the SSM and stellar evolution.  The
arguments for measuring the CNO flux, using our sun as a calibrated laboratory, seem
particularly strong.  Such a program would effectively test our understanding of the
hydrogen burning mechanism for massive main-sequence stars.  It would also address
the primary discrepancy in the SSM, the tension between helioseismology and 
neutrino flux predictions that follows from new analyses of surface metallicity.

\end{document}